\documentclass[reprint,prd,twocolumn,nofootinbib,letter,10pt]{revtex4-1}
\usepackage{graphicx,epsfig,fancyvrb,makeidx,latexsym,amssymb,amsmath,color}

\allowdisplaybreaks

\usepackage[colorlinks=true,linkcolor=blue,citecolor=blue]{hyperref}

\definecolor{greeen}{rgb}{0.03,0.84,0.13}
\definecolor{test}{rgb}{0.03,0.74,0.33}
\definecolor{viol}{rgb}{0.44,0,0.94}
\definecolor{or}{rgb}{0.95,0.65,0}

\begin{document}

\title{Prospects of type-II seesaw at future colliders in light of the DAMPE $e^+ e^-$ excess}



\author{Yicong Sui}
\email{yicongsui@wustl.edu}
\affiliation{Department of Physics and McDonnell Center for the Space Sciences,  Washington University, St. Louis, MO 63130, USA}

\author{Yongchao Zhang}
\email{yongchao.zhang@physics.wustl.edu}
\affiliation{Department of Physics and McDonnell Center for the Space Sciences,  Washington University, St. Louis, MO 63130, USA}

\begin{abstract}
  The DAMPE $e^+ e^-$ excess at around 1.4 TeV could be explained in the type-II seesaw model with a scalar dark mater $D$ which is stabilized by a discrete $Z_2$ symmetry. The simplest scenario is the annihilation $DD \to H^{++} H^{--}$ followed by the subsequent decay $H^{\pm\pm} \to e^\pm e^\pm$, with both the DM and triplet scalars roughly 3 TeV with a small mass splitting. In addition to the Drell-Yan process at future 100 TeV hadron colliders, the doubly-charged components could also be produced at lepton colliders like ILC and CLIC in the off-shell mode, and mediate lepton flavor violating processes $e^+ e^- \to \ell_i^\pm \ell_j^\mp$ (with $i \neq j$). A wide range of parameter space of the type-II seesaw could be probed, which are well below the current stringent lepton flavor constraints.
\end{abstract}

\maketitle

\section{Introduction}

The dark matter (DM) constitutes about 26\% of the energy budget of the Universe, and has been well established from astrophysical and cosmological observations~\cite{Ade:2015xua}, though its existence has yet to be confirmed by the direct detection experiments, as well as on the high energy colliders or via the indirect detection experiments. Very recently, the DArk Matter Particle Explorer (DAMPE) collaboration~\cite{TheDAMPE:2017dtc} has found a sharp peak at around 1.4 TeV in the cosmic ray $e^+ e^-$ spectrum~\cite{Ambrosi:2017wek}, which might be challenging to be understood in terms of the astrophysical source,  but can be interpreted by the annihilation or decay of DM particles at the TeV scale~\cite{Yuan:2017ysv}, like the intensive phenomenological studies in Refs.~\cite{Fan:2017sor, Gu:2017gle, Duan:2017pkq, Zu:2017dzm, Tang:2017lfb, Chao:2017yjg, Gu:2017bdw, Athron:2017drj, Cao:2017ydw, Duan:2017qwj, Liu:2017rgs, Huang:2017egk, Chao:2017emq, Gao:2017pym, Niu:2017hqe, Gu:2017lir, Nomura:2017ohi, Zhu:2017tvk, Ghorbani:2017cey, Cao:2017sju, Li:2017tmd, Chen:2017tva,  Jin:2017qcv, Cholis:2017ccs, Fang:2017tvj, Ding:2017jdr, Yang:2017cjm, Ge:2017xyz, Liu:2017abc, Zhao:2017nrt}. It is widely believed that the DM particle is preferably to be leptophilic and the annihilation cross section ${\rm DM} \, {\rm DM} \to e^+ e^-$ (or ${\rm DM} \, {\rm DM} \to XX \to e^+ e^+ e^- e^-$, with $X$ playing the role of ``mediator'') is orders of magnitude higher than that for the freezing out of thermal DM in the early universe. One can assume a nearly subhalo structure~\cite{Yuan:2017ysv} or use Sommerfeld enhancement~\cite{Feng:2009hw, Feng:2010zp} to obtain such a large ``boost factor''.

In this paper we propose a well-motivated neutrino model to explain he DAMPE excess data, based on the minimal type-II seesaw mechanism~\cite{type2a,type2b,type2c,type2d,type2e}. To understand the tiny neutrino masses, an isospin triplet scalar is added to the SM, which is automatically leptophilic in the sense that all the neutral, singly and doubly charged components of the triplet decays predominantly into the SM leptons in a large parameter space. A scalar DM $D$ is introduced to the minimal type-II seesaw model, which is stabilized by a discrete $Z_2$ symmetry~\cite{Gogoladze:2009gi,Dev:2013hka}. Then an economical explanation of the $e^+ e^-$ excess at 1.4 TeV could be the annihilation of DM with mass $\sim 3$ TeV into the doubly-charged scalars which decays further into $e^\pm e^\pm$ pairs, i.e. 
$DD \to H^{++} H^{--} \to e^+ e^+ e^- e^-$.
Note that the annihilation process here is even simpler than in Refs.~\cite{Dev:2013hka, Li:2017tmd,Ding:2017jdr}, where the authors considered also the annihilation of DM into the neutral and singly-charged scalars from the triplet. The decay of these scalars gives rise to TeV-scale, almost monochromatic neutrinos, which might be constrained by the IceCube data~\cite{Zhao:2017nrt,Aartsen:2016xlq}.

As the fitting of DAMPE data has been done in Ref.~\cite{Li:2017tmd,Ding:2017jdr}, in this work we concentrate more on the phenomenological implication of the existence of a 3 TeV DM in the type-II seesaw model, and the prospects and testability at future lepton such as ILC~\cite{Baer:2013cma} and CLIC~\cite{Battaglia:2004mw} and the 100 TeV hadron colliders like FCC-hh~\cite{Golling:2016gvc, Contino:2016spe} and SPPC~\cite{CEPC-SPPCStudyGroup:2015csa}. By interacting with the SM Higgs, all the triplet components can be pair produced in the scalar portal. The cross sections for the singly and doubly charged scalars are too small when compared to the Drell-Yan processes; while the neutral components decays predominantly into neutrinos, they can, in principle, be limited by the monojet searches of DM at hadron colliders~\cite{Khachatryan:2014rra, Aad:2015zva, Aaboud:2016tnv, Xiang:2015lfa,Malik:2014ggr}.
The doubly-charged scalars could induce lepton flavor violating (LFV) processes at future lepton colliders, e.g. $e^+ e^- \to \ell_i^\pm \ell_j^\mp$ (with the flavor indices $i \neq j$)~\cite{Dev:2017ftk}, though they can not be (pair) produced on-shell. It turns out that a large region in the parameter space of type-II seesaw could be probed in these LFV channels, that are well below the current lepton flavor limits like $\ell_i \to \ell_j \gamma$, $\ell_i \to 3\ell_j$, $\ell_i \to 2\ell_j \ell_k$ and the anomalous magnetic moments of electron and muon~\cite{PDG}.

The rest of this paper is organized as follows: The minimal type-II seesaw is sketched in Section~\ref{sec:model}, and Section~\ref{sec:dm} is devoted to the DM phenomenology, followed by the fitting of DAMPE excess in Section~\ref{sec:excess}. The hadron and lepton collider prospects are presented in Section~\ref{sec:collider}, before we conclude in Section~\ref{sec:conclusion}.


\section{Type-II seesaw model}
\label{sec:model}
In the type-II seesaw model~\cite{type2a,type2b,type2c,type2d,type2e}, an isospin triplet is added to the scalar sector, which has a hypercharge of $2$, i.e.
\begin{eqnarray}
\Delta \ = \ \left(\begin{array}{cc}
\Delta^+/\sqrt 2 & \Delta^{++}\\
\Delta^0 & -\Delta^+/\sqrt 2
\end{array}
\right) \,.
\end{eqnarray}
Following the notation in~\cite{DGOS}, the most general scalar potential for the SM doublet $\phi$ and the triplet $\Delta$ reads
\begin{eqnarray}
{\cal V}(\phi,\Delta) &=&
-\mu_\phi^2(\phi^\dag \phi)
+\mu^2_\Delta {\rm Tr}({\Delta} ^\dag {\Delta}) \nonumber \\
&& +\frac{\lambda}{2}(\phi^\dag \phi)^2
+ \frac{\lambda_1}{2}\left[{\rm Tr}({\Delta}^\dag {\Delta})\right]^2\nonumber\\
&&
+\frac{\lambda_2}{2}\left(\left[{\rm Tr}({\Delta} ^\dag {\Delta})\right]^2
-{\rm Tr}\left[({\Delta} ^\dag {\Delta})^2\right]\right) \nonumber \\ && +\lambda_4(\phi^\dag \phi){\rm Tr}({\Delta} ^\dag {\Delta})+\lambda_5\phi^\dag[{\Delta}^\dag,{\Delta}]\phi\nonumber\\
&&
+\left(\frac{\lambda_6}{\sqrt 2}\phi^{\sf T}i\sigma_2{\Delta}^\dag \phi+{\rm H.c.}\right)\, ,
\label{eq:Vpd}
\end{eqnarray}
with all the couplings being real. A non-zero vacuum expectation value (VEV) for the Higgs doublet field $\langle \phi^0 \rangle = v_{\rm EW}/\sqrt2$ (with $v_{\rm EW} \simeq$ 246 GeV) induces a tadpole term for the scalar triplet field ${\Delta}$ via the $\lambda_6$ term in Eq.~(\ref{eq:Vpd}), thereby generating a non-zero VEV for its neutral component, $\langle \delta^0\rangle = v_\Delta/\sqrt 2$, and breaking lepton number by two units.

As the VEV $v_\Delta$ is in charge of the tiny neutrino masses, it is expected to be much smaller than the electroweak scale, or even close to the eV scale. In the limit of $v_\Delta \ll v_{\rm EW}$, after spontaneous symmetry breaking, the neutral, singly-charged and doubly-charged components of the triplet obtain their masses,
\begin{eqnarray}
m^2_{H,\,A} &=& \mu_\Delta^2+\frac{1}{2}(\lambda_4-\lambda_5)v_{\rm EW}^2 ,
\label{mass3} \\
m^2_{H^\pm} &=& \mu_\Delta^2+\frac{1}{2}\lambda_4v_{\rm EW}^2 \,,
\label{mass2}\\
m^2_{H^{\pm\pm}} &=& \mu_\Delta^2+\frac{1}{2}(\lambda_4+\lambda_5)v_{\rm EW}^2 \,,
\label{mass1}
\end{eqnarray}
with the neutral component from the doublet has a mass $m_h^2=\lambda v_{\rm EW}^2$ and identified as the SM Higgs. As a direct result of $v_\Delta \ll v_{\rm EW}$, the mixing of doublet and triplet components are generally very small. however, that does {\it not} necessarily mean their couplings are also very small, much like the couplings of right-handed triplet to the SM Higgs in the left-right symmetric model~\cite{Dev:2016dja}. In particular, the  couplings $\lambda_{4,5}$ might be large, say order one, if these heavy scalars are at the TeV scale, as implied by the DAMPE data, and this would induce the pair production of the triplet scalars in the SM Higgs portal,  which is largely complementary to the gauge portal like the Drell-Yan process; see Section~\ref{sec:collider} and Ref.~\cite{nextpaper} for more details.

The triplet $\Delta$ couples to the SM lepton doublet  $L=(\nu,\ell)_L^{\sf T}$ via the Yukawa interactions
\begin{eqnarray}
{\cal L}_Y &=& -\frac{1}{\sqrt 2}\left(Y_\Delta\right)_{ij} L_i^{\sf T}Ci\sigma_2 {\Delta} L_j+{\rm H.c.},
\label{lag}
\end{eqnarray}
with $C$ the charge conjugation matrix. Then the tiny neutrino mass matrix is obtained with the induced VEV $v_\Delta$:
\begin{eqnarray}
(m_\nu)_{ij} = v_\Delta (Y_\Delta)_{ij}
= U^{\sf T}\widehat{m}_\nu U \, .
\label{eq:neutrino}
\end{eqnarray}
The Yukawa coupling matrix $Y_\Delta$ is completely fixed by the observed neutrino mass squared differences and mixing angles, with $\widehat{m}_\nu$ the diagonal neutrino masses and $U$ the standard PMNS matrix, once the lightest neutrino mass is known. Under the condition
\begin{eqnarray}
Y_\Delta \sim \frac{m_\nu}{v_{\Delta}} \gg \frac{v_{\Delta}}{v_{\rm EW}} \,,
\end{eqnarray}
i.e. $v_\Delta \ll 0.1$ MeV, the couplings of triplet scalars to the leptons are much larger than those to the SM Higgs, and the triplet scalars decays predominantly into the SM charged leptons and neutrinos~\cite{Perez:2008ha}:\footnote{Note that if the type-II seesaw is embedded in the left-right framework, the neutral components of the right-handed triplet have much richer decay modes as they could have sizable mixing with the SM Higgs and couple to the extra gauge bosons~\cite{Dev:2016dja}.}
\begin{eqnarray}
H,\,A \to \nu_i \bar{\nu}_j \,,\;\;
H^\pm \to \ell^\pm_i \nu_j (\bar{\nu}_j) \,,\;\;
H^{\pm\pm} \to \ell_i^{\pm} \ell_j^{\pm} \,.
\end{eqnarray}
If the triplet scalars are well above the electroweak scale, as implied by the DAMPE data, such a low-scale $v_\Delta$ is safe from the current constraints of the electroweak precision data~\cite{PDG}, LFV processes such as $\ell_i \to \ell_j \gamma$ and $\ell_i \to 2\ell_j \ell_k$~\cite{lfv1, lfv2}.

With the following latest neutrino oscillation data~\cite{PDG}
\begin{eqnarray}
&&
\Delta m^2_{\rm sol}=  7.6\times 10^{-5}~{\rm eV}^2,~
\Delta m^2_{\rm atm} = 2.4\times 10^{-3}~{\rm eV}^2,~\nonumber\\
&&
\theta_{12}=34^\circ,~ \theta_{23}=45^\circ,\theta_{13}=8.8^\circ
\end{eqnarray}
and the preferred Dirac CP violating phase $\delta = 3\pi/2$~\cite{Abe:2015awa}, we can predict the flavor contents of the decay products of the leptophilic triplet scalars. Summing over all the flavor conserving and violating decays, the flavor fractions are expected to be
\begin{eqnarray}
\label{fractions}
\text{Normal hierarchy}:\ &&e:\mu:\tau \ = \ 0.032: 0.484 : 0.484 \,, \nonumber \\
\text{Inverted hierarchy}:\ && e:\mu:\tau \ = \ 0.48: 0.26 : 0.26
\end{eqnarray}
for the two mass orderings, in the limit of massless lightest neutrino. In the degenerate neutrino mass limit, both the two orderings approach to be $1:1:1$, which is however highly disfavored by the current cosmological constraints on neutrino masses $\sum_i m_i < 0.23$ eV at the 95\% CL~\cite{Planck}.

\section{Dark Matter}
\label{sec:dm}
The minimal type-II seesaw model can be extended to accommodate a cold DM candidate by simply adding a SM singlet real scalar field $D$~\cite{Gogoladze:2009gi}. Its stability can be ensured by assigning it an odd $Z_2$-parity, whereas all other fields are even under the $Z_2$ symmetry. The scalar potential relevant for the DM physics is given by
\begin{eqnarray}
{\cal V}_{\rm DM} &=& \frac{1}{2}\mu_D^2D^2+\lambda_D D^4 \nonumber \\
&& +\lambda_\phi D^2(\phi^\dag \phi)
+\lambda_\Delta D^2{\rm Tr}({\Delta}^\dag {\Delta}) \,,
\label{dm}
\end{eqnarray}
Then the DM mass is given by $m_{D}^2=\mu_D^2+\lambda_\phi v_{\rm EW}^2+\lambda_\Delta v_\Delta^2$ after spontaneous symmetry breaking of both the doublet and triplet.

One could have a fermionic DM instead, and one well-motivated example is the embedding of type-II seesaw in the left-right framework based on the gauge group $SU(3)_C \times SU(2)_L \times SU(2)_R \times U(1)_{Y_L} \times U(1)_{Y_R}$~\cite{Dev:2016qbd,Dev:2016xcp,Dev:2016qeb}, where the lightest right-handed neutrino (RHN) $N$ in the heavy sector is stabilized by an automatic $Z_2$ symmetry, which is the residual lepton number in the $SU(2)_R$ sector. In light of the DAMPE data, the RHN DM could annihilate into $e^+ e^-$ through both the gauge and scalar portals, which, however, are both suppressed by either the small electron mass or the small DM velocity ($p$-wave mode). This makes the left-right framework less attractive and we focus here only on the scalar DM $D$.

Given the quartic couplings $\lambda_\Delta$ and $\lambda_\phi$ in Eq.~(\ref{dm}), the DM $D$ could annihilate directly into scalar pairs in the doublet and triplet, i.e. $DD\to hh,HH,AA, H^{+}H^{-}, H^{++}H^{--}$, with the leading order thermalized cross section, in the non-relativistic limit,
\begin{eqnarray}
\langle \sigma v\rangle = \frac{1}{16\pi m_{D}^2}\left[\lambda_\phi^2\sqrt{1-\frac{m_h^2}{m_{D}^2}}
+6\lambda_\Delta^2\sqrt{1-\frac{m_\Delta^2}{m_{D}^2}}\right] \,.
\label{an}
\end{eqnarray}
The factor of $6$ in the second term counts all the degrees of freedom in the triplet sector assuming they are (almost) mass degenerate $m_{H,A} = m_{H^\pm} = m_{H^{\pm\pm}} = m_\Delta$. With a small coupling $\lambda_\phi$, the proton/antiproton flux from DM annihilation $DD \to hh$ will be suppressed. In addition, with a sufficiently small $\lambda_\phi$, the spin independent scattering of DM off the nuclei in direct direction experiments, which is mediated by an SM Higgs in the $t$-channel, could be safely below the current limits from LUX~\cite{Akerib:2016vxi}, Xenon1T~\cite{Aprile:2017iyp} and PandaX experiment~\cite{Cui:2017nnn}. With the singly and doubly charged scalars decaying further into charged leptons (and neutrinos), we can explain the DAMPE $e^+ e^-$ data via $D D \to H^+ H^-,\, H^{++} H^{--}$, as long as the DM mass $m_{D} \simeq 3$ TeV and the charged scalar masses are slightly below $m_{D}$ to have nearly monochromatic $e^\pm$ from the scalar decay.

In light of the mass quasi-degeneracy of triplet scalars and their universal coupling $\lambda_\Delta$ to the DM $D$, a sizable portion of DM particles annihilate into the neutral and singly-charged scalars which decay further into TeV scale primary neutrinos (and charged leptons). This might be tightly constrained by the IceCube neutrino data~\cite{Zhao:2017nrt, Aartsen:2016xlq}, can could be easily evaded when the small  mass splitting of the triplet scalars is taken into consideration, cf. Eqs.~(\ref{mass3}) to (\ref{mass1}), depending on the $\lambda_5$ parameter in the limit of $v_\Delta \ll v_{\rm EW}$. Setting the overall scale of $m_\Delta = 3$ TeV in light of the DAMPE data, a $\lambda_5$ of ${\cal O} (1)$ leads to a splitting of order 10 GeV:\footnote{Though there might be unitarity and stability constraints on the quartic coupling $\lambda_5$~\cite{Dev:2017ouk}, it would however be weakened to some extent by the existence of the DM scalar $D$ and its interactions with the triplet in Eq.~(\ref{dm}).}
\begin{eqnarray}
\label{eqn:splitting}
&& m_{H^\pm} - m_{H^{\pm\pm}} = \frac12 (m_{H,A} - m_{H^{\pm\pm}}) \nonumber \\
&& = -\frac{\lambda_5 v_{\rm EW}^2}{4 m_\Delta}
\simeq - \lambda_5 \times (5 \, {\rm GeV}) \,.
\end{eqnarray}
Then with a negative $\lambda_5$, we can have the mass ordering of the DM and the triplet scalars
\begin{eqnarray}
m_{H^{\pm\pm}} < m_{D} < m_{H^\pm} < m_{H,A} \,.
\end{eqnarray}
This renders that the DM pairs annihilate only predominately into the doubly-charged scalars, thus evading the potential neutrino constraints from the decays of neutral and singly-charged scalars~\cite{Aartsen:2016xlq}. Then the annihilation cross section in Eq.~(\ref{an}) can be slightly simplified
\begin{eqnarray}
\label{eqn:sigmav}
\langle \sigma v\rangle = \frac{\lambda_\Delta^2}{8\pi m_{D}^2}
\sqrt{1-\frac{m_{H^{\pm\pm}}^2}{m_{D}^2}} \,,
\end{eqnarray}
which could easily produce the observed DM density of $\Omega_{\rm DM} h^2 \simeq 0.12$ via~\cite{Kolb:1990vq}
\begin{eqnarray}
\label{eqn:relicdensity}
\Omega_D h^2 = \frac{1.07 \times 10^{9} \, {\rm GeV}^{-1}}{M_{\rm Pl}}
\frac{x_F}{\sqrt{g_\ast}} \langle \sigma v \rangle^{-1}
\end{eqnarray}
if $\lambda_\Delta = 2.0 \, (3.5)$ for $m_{D} = 3$ TeV and $m_{D} - m_{H^{\pm\pm}} = 10 \, (1)$ GeV. In the equation above $M_{\rm Pl}$ is the Planck scale, $x_F = m_D /T_F \simeq 20$ (with $T_F$ being the freeze-out temperature), $g_\ast = 106.75$ the relativistic degrees of freedom at $T_F$.

\begin{figure}[!t]
  \centering
  \includegraphics[width=0.45\textwidth]{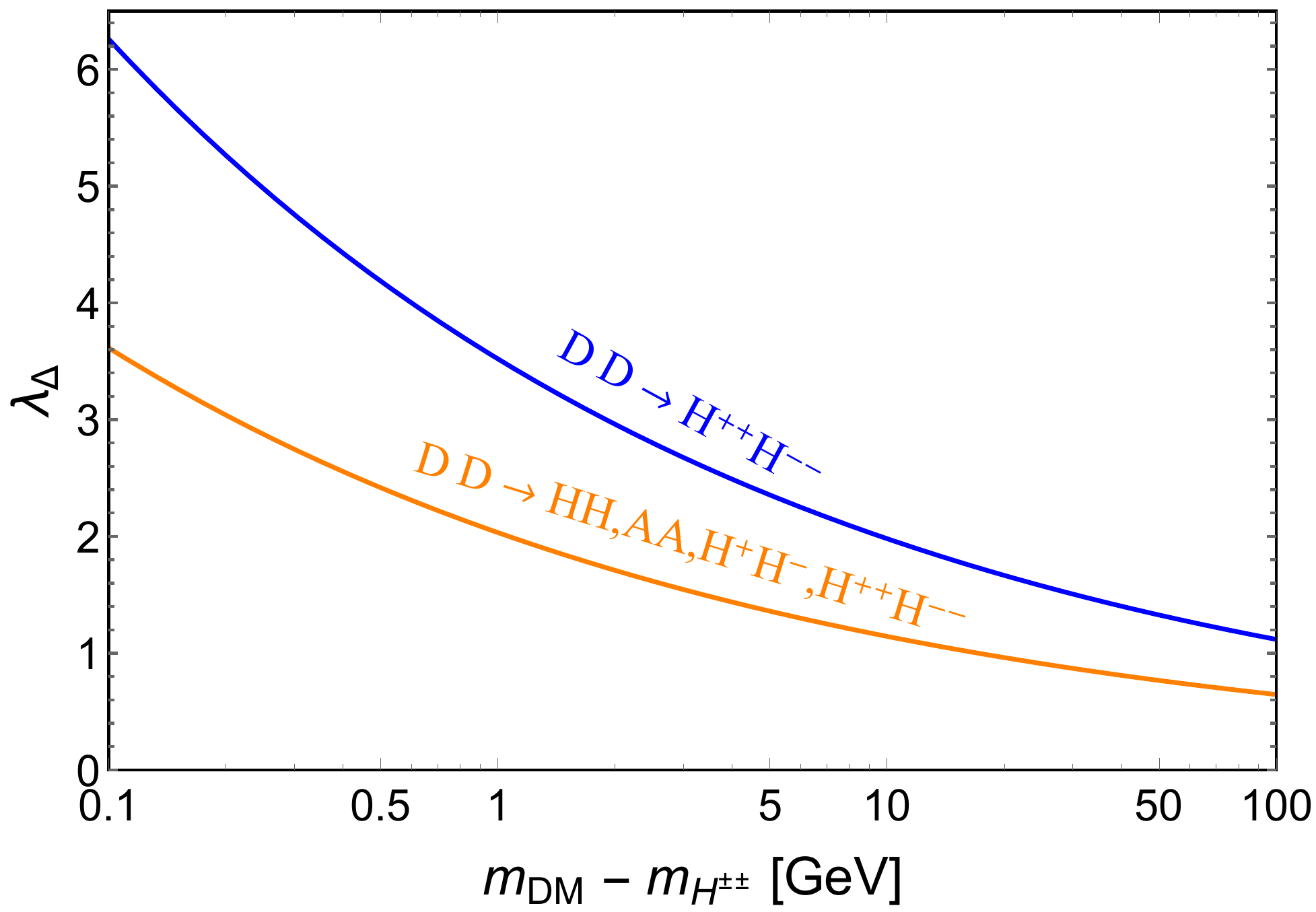}
  \caption{Dependence of the quartic coupling $\lambda_{\Delta}$ on the mass splitting $m_{D} - m_{H^{\pm\pm}}$ for the two annihilation case as indicated. Along these two curves we can obtain the observed DM relic density via Eqs.~(\ref{eqn:sigmav}) and (\ref{eqn:relicdensity}). }
  \label{fig:parameter}
\end{figure}

More generic dependence of the quartic coupling $\lambda_\Delta$ on the mass splitting is presented in Fig.~\ref{fig:parameter}, where we have set explicitly $\lambda_\phi = 0$. For the sake of comparison, we show both the two curves for respectively $DD \to H^{++} H^{--}$ and $DD \to HH,AA,H^+ H^-,H^{++} H^{--}$, with all the scalars mass degenerate (the splitting in Eq.~(\ref{eqn:splitting}) is zero, i.e. $\lambda_5$ = 0). It is transparent in Figure~\ref{fig:parameter} that though in the simplified case of $DD \to H^{++} H^{--}$ the quartic couplings $\lambda_\Delta$ is required to be larger, it is still within the perturbative limit of $4\pi$ for a small splitting of order 0.1 GeV, such that we can have the observed relic density of cold DM.

\section{DAMPE $e^+ e^-$ excess}
\label{sec:excess}

To fit the DAMPE excess, we adopt the simplest scenario above, i.e.
\begin{eqnarray}
DD \to H^{++} H^{--} \to e^+ e^+ e^- e^- \,,
\end{eqnarray}
with a branching ratio of ${\rm BR} (H^{\pm\pm} \to e^\pm e^\pm) = 3.2\% \, (48\%)$ for the normal (inverted) hierarchy in the massless lightest neutrino limit (cf. Eq.~(\ref{fractions})). To be specific, we set explicitly the DM mass $m_{D} = 3$ TeV and the small mass splitting $m_{D} - m_{\rm H^{\pm\pm}} = 10$ GeV. Then the electrons from doubly-charged scalar decay are almost monochromatic, with an energy width of 10 GeV at the source, i.e. the energy distribution
\begin{eqnarray}
\frac{d\, N}{d\,E} \simeq
\begin{cases}
0.1 \, {\rm GeV}^{-1} \,, & E_{e^\pm} \in [1.49,\, 1.5] \, {\rm TeV} \\
0 \,, & {\rm otherwise} \,.
\end{cases}
\end{eqnarray}

To have a large ``boost factor'' of order $10^2$ for the DAMPE  data, we assume there is a DM subhalo at the distance of 0.3 kpc with a local density of $\rho_s = 100 \, {\rm GeV}/{\rm cm}^3$, with the standard NFW density profile~\cite{Navarro:1996gj}
\begin{eqnarray}
\frac{\rho_{D} (r)}{\rho_s}  = \frac{(r/r_s)^{-\gamma}}{(1+r/r_s)^{3-\gamma}} \,,
\end{eqnarray}
with $r_s = 0.1$ kpc and $\gamma = 0.5$~\cite{Simon:2007dq,Walker:2008ax}.

The propagation of $e^\pm$ from  the subhalo source to the Earth is estimated by solving the diffuson equation
\begin{equation}
\partial_t f - \partial_E (b(E) f) - D(E) \nabla^2 f \ = \ Q \,,
\end{equation}
with $f$ the electron energy spectrum and $Q$ the source term. $b(E)=b_0 (E/\text{GeV})^2$ is the energy loss coefficient with $b_0=10^{-16}$ GeV/s, and $D(E)=D_0(E/\text{GeV})^\delta$ is the diffusion coefficient, with $D_0$ = 11 pc$^2$/kyr, and $\delta=0.7$~\cite{Cirelli:2008id}. The general solution to the diffusion equation above can be written in the form below, for the steady-state case~\cite{Kuhlen:2009is,Delahaye:2010ji},
\begin{eqnarray}
&& f({\bf x}, E) = 4\int d^3 x_s \int dE_s\,
G({\bf x}, E; {\bf x}_s, E_s) Q({\bf x}_s, E_s) \,, \nonumber \\ &&
\end{eqnarray}
where the factor of $4$ counting the numbers of $e^\pm$ from a pair of DM annihilation, the source
\begin{eqnarray}
Q({\bf x}, E) = {1 \over 2} { \rho^2_{D}({\bf x}) \over m_{D}^2 }
\langle \sigma v \rangle {dN \over dE} \,,
\end{eqnarray}
the Green function
\begin{eqnarray}
G({\bf x}, E; {\bf x}_s, E_s) = {\exp\left[-{({\bf x-x}_s)^2/\lambda^2}\right]
\over b(E) (\pi \lambda^2)^{3/2}} \,,
\end{eqnarray}
with the propagation scale
\begin{eqnarray}
\lambda = 2 \left[ \int_E^{E_s} dE' {D(E') \over b(E')} \right]^{1/2} \,.
\end{eqnarray}
Then the $e^\pm$ flux is given by $\Phi({\bf x}, E) = v f({\bf x}, E)/(4\pi)$, with $v \simeq c$ the velocity of $e^\pm$.

\begin{figure}[!t]
  \centering
  \includegraphics[width=0.48\textwidth]{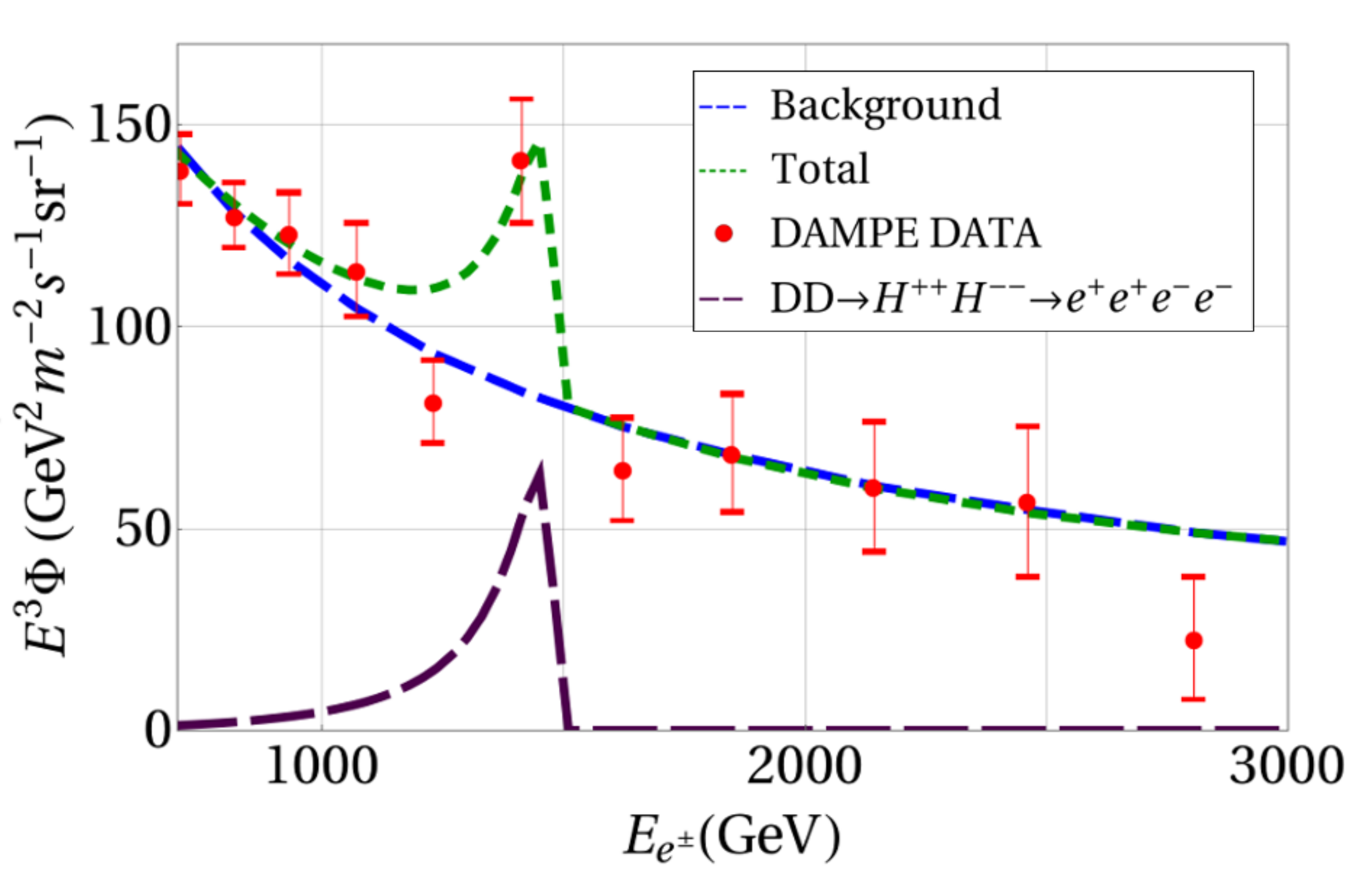}
  \caption{Fitting of the DAMPE data, with the power law background, the $e^\pm$ from DM annihilation, the total fitting data and the DAMPE data. }
  \label{fig:fitting}
\end{figure}

The fitting of the DAMPE data with the inverted hierarchy of neutrinos is presented in Fig.~\ref{fig:fitting}, where we have also shown the power law background and the DAMPE data. It is transparent that a $\sim3$ TeV DM could easily explain the peak in the $e^+ e^-$ spectrum. If the neutrinos are of normal hierarchy, then the ``boost'' factor is required to be 15 times larger and the muon and tauon decays give rise to softer secondary electrons/positrons, which is less favored by the DAMPE data~\cite{Ding:2017jdr}. To be consistent, we have checked also the photons~\cite{Ackermann:2014usa} and neutrinos~\cite{Aartsen:2016xlq} from the decays of $H^{\pm\pm} \to \ell_i^\pm \ell_j^\pm$ involving muon and tauons, following~\cite{Chianese:2016kpu, Murase:2014tsa}. As expected, these secondary particles are much softer than the electrons at 1.4 TeV, and the fluxes are orders of magnitude below the current astrophysical backgrounds.

\section{Prospects at future colliders}
\label{sec:collider}

Inspired by the tentative DAMPE excess at 1.4 TeV, one of the most important implications for the minimal type-II seesaw model is that the triplet resides at the 3 TeV scale. That is too heavy to be directly pair produced at the LHC or the ILC running at 1 TeV, and might only be probable on-shell at future 100 TeV hadron colliders~\cite{Arkani-Hamed:2015vfh}. In this section we scrutinize how the TeV scale scalars could be tested at future lepton and hadron colliders, in particular via searches of LFV processes $e^+ e^- \to \ell_i^\pm \ell_j^\mp$ at lepton colliders, which is mediated by an off-shell doubly-charged scalar~\cite{Dev:2017ftk}.

\subsection{Prospects at 100 TeV hadron collider}

In the DM sector, their is only one term that connects the scalar DM to the SM particles, i.e. the $\lambda_\phi$ term in Eq.~(\ref{dm}). Given this term, the SM Higgs mediates the spin independent scattering of DM off the nuclei in direct detection experiments, with the cross section~\cite{Li:2017tmd} $\sigma_{\rm SI} = 3 \times 10^{-9} \, {\rm pb} \times \lambda_{\phi}^2$, which requires that $|\lambda_\phi| < 0.95$ for a DM mass of 3 TeV. The coupling $\phi$ can also induce the annihilation of DM into the SM Higgs pairs, i.e. $DD \to hh$, which decays further into $b \bar{b}$, $WW$, $ZZ$ etc, giving rise to antiprotons over the cosmological background~\cite{Adriani:2008zq, Adriani:2010rc, Aguilar:2016kjl}. It turns out that the antiproton limits are more stringent, $\lambda_{\phi} \gtrsim 0.06$~\cite{Dev:2013hka}.

This could be further cross checked at high energy hadron collider, e.g. via the process
\begin{eqnarray}
\label{eqn:DMcollider}
gg \to h^\ast \, g \to D D g \,.
\end{eqnarray}
generating the monojet plus missing $E_T$ signal. However, this is highly suppressed by the DM mass of $2m_D = 6 \, {\rm TeV}$: even at future 100 TeV collider, the cross section is only $10^{-3} \, {\rm fb} \times \lambda_\Phi^2$ when we adopt a $p_T$ cut of 500 GeV on the primary jet using {\tt CalcHEP}~\cite{Belyaev:2012qa}. It is almost impossible to set any limits on the coupling $\lambda_\phi$ from the monojet searches.



Even if the coupling $\lambda_\phi$ is small, the model could still be tested via the the searches of triplet scalars at future hadron and lepton colliders, which does not only play the role of ``mediator'' connecting the SM and DM sectors but also be responsible for the neutrino mass generation. It is worth pointing out that in addition to the gauge portal i.e. the Drell-Yan processes, the triplet scalars could also be produced in the scalar portal, i.e. couplings to the SM Higgs, in particular for the neutral components $H$ and $A$.

To be specific, the couplings for  $h HH$ ($hAA$), $h H^+ H^-$ $h H^{++} H^{--}$ are respectively $(\lambda_4 - \lambda_5) v_{\rm EW}$, $\lambda_{4} v_{\rm EW}$ and $(\lambda_4 + \lambda_5) v_{\rm EW}$ (cf. Eqs.~(\ref{mass3}) to (\ref{mass1})), although the couplings $H hh$ and $Ahh$ are expected to be small, highly suppressed by the VEV $v_\Delta$.  Then the CP-even and odd components $H$ and $A$ can be pair produced from the SM Higgs through gluon fusion, in association with a gluon jet; after produced, these heavy scalars decay predominantly into neutrinos, i.e.\footnote{For larger VEV $v_\Delta \gtrsim 0.1$ MeV, $H$ and $A$ could decay into the SM particles like $WW$ through mixing with the SM Higgs, with a sizable branching fraction, e.g. $gg \to h^\ast \to HH \to 4W$. It is even possible to have displaced jets/leptons from the subsequent decays $W \to jj,\, \ell\nu$~\cite{Perez:2008ha}.}
\begin{eqnarray}
\label{eqn:monojet}
gg \to h^\ast g \to (HH/AA)g \to \nu\nu \bar\nu \bar\nu \, g \,.
\end{eqnarray}
At hadron collider, this turns out to be missing transverse energy plus jet(s), much like the ``real'' DM process in Eq.~(\ref{eqn:DMcollider}). However, this is also highly suppressed by the large scalar mass $m_{H,A} \simeq m_D \simeq 3$ TeV: For the benchmark value of $(\lambda_4 - \lambda_5) = 1$ and the cut $p_T > 500$ GeV on the primary jet~\cite{Khachatryan:2014rra, Aad:2015zva, Aaboud:2016tnv, Xiang:2015lfa,Malik:2014ggr}, the total cross section for process in Eq.~(\ref{eqn:monojet}) is only $\simeq 7 \times 10^{-4}$ fb at a 100 TeV collider, and thus it is very challenging to be tested at hadron colliders in the monojet channel.


The singly and doubly-charged scalars $H^\pm$ and $H^{\pm\pm}$ could also be produced in the scalar portal, i.e.
\begin{eqnarray}
gg \to h^\ast \to H^+ H^- / H^{++} H^{--} \,,
\end{eqnarray}
and decay into charged leptons (and neutrinos) with potential lepton flavor violating signals $H^{++}H^{--} \to \ell_i^+ \ell^+_j \ell_m^- \ell^-_n$ and $H^+ H^- \to \ell_i^+ \ell_j^- \nu\bar{\nu}$. This is largely complementary to the Drell-Yan production in the gauge portal. For $m_\Delta \simeq 3$ TeV, the production cross sections for the Drell-Yan processes of the singly and doubly charged scalars at 100 TeV hadron collider are respectively 0.016 fb and 0.064 fb~\cite{Arkani-Hamed:2015vfh}; in the scalar portal they are much smaller, being both $3.8 \times 10^{-4} \, {\rm fb}$~\cite{Dev:2016dja}, up to the couplings squared ($\lambda_4^2$ and $(\lambda_4 + \lambda_5)^2$).

\begin{figure}[!t]
  \centering
  \includegraphics[width=0.48\textwidth]{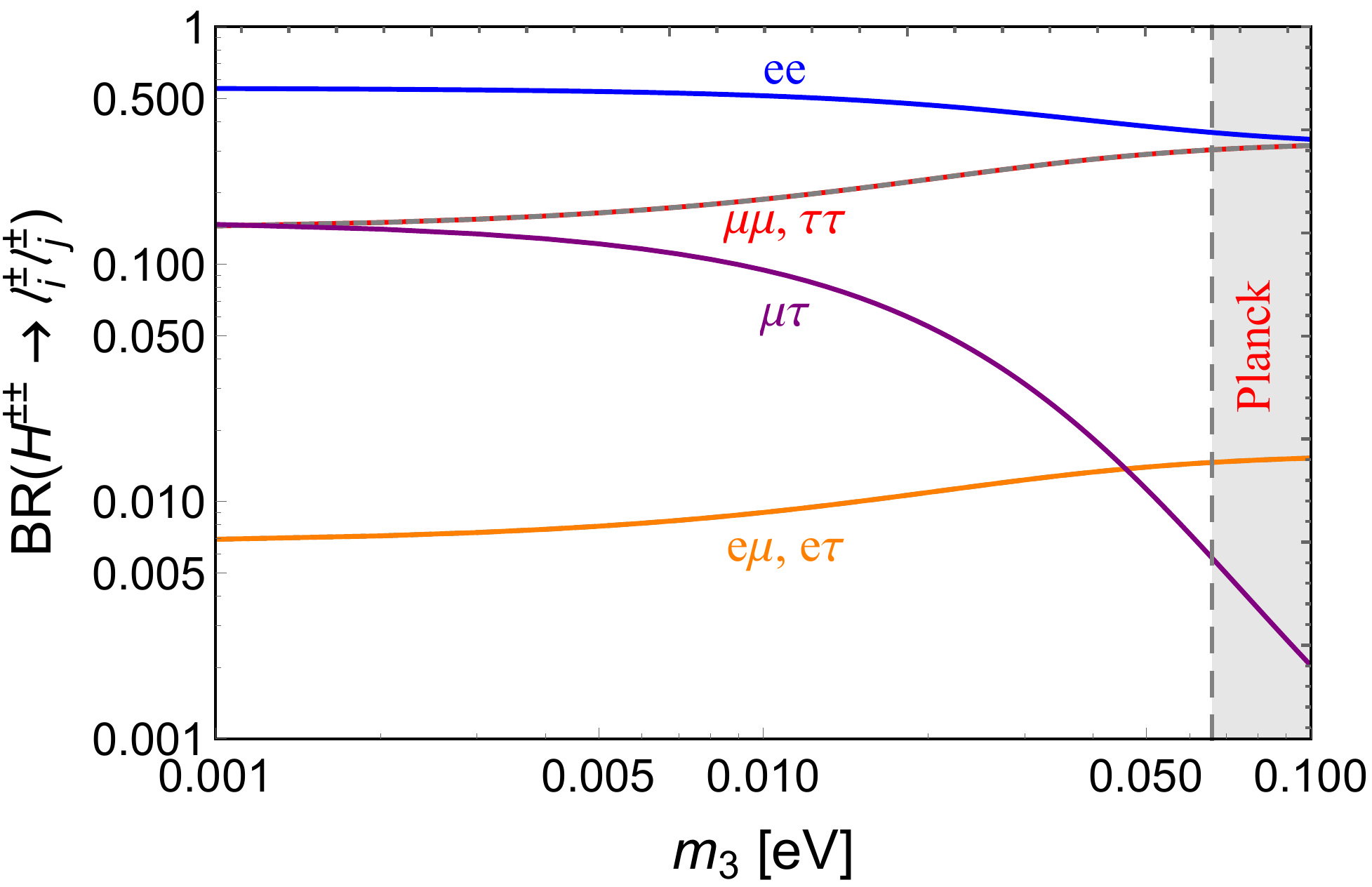}
  \caption{Decay branching ratio of the doubly charged scalar into the charged leptons ${\rm BR} (H^{\pm\pm} \to \ell_i^\pm \ell_j^\pm)$ as functions of the lightest neutrino mass $m_3$ in the inverted hierarchy. The gray region on the right is excluded by the Planck data~\cite{Planck}.}
  \label{fig:BR}
\end{figure}

Once the lightest neutrino mass ($m_3$ in the inverted hierarchy) is known, the branching fractions of ${\rm BR} (H^{\pm\pm} \to \ell_i^\pm \ell_j^\pm)$ can be completely determined via the neutrino mass matrix Eq.~(\ref{eq:neutrino}), as shown in Fig.~\ref{fig:BR}~\cite{Perez:2008ha}. We can read from the figure that the most promising signal is the lepton number violating (LNV) decays like $(e^+ e^+) (e^- e^-)$ and $(e^+ e^+) (\mu^- \mu^-)$ with resonance of the same-sign leptons $m_{\ell\ell'} \simeq m_{H^{\pm\pm}} \simeq 3$ TeV. The primary LFV search would be
\begin{eqnarray}
pp \to H^{++} H^{--} \to e^\pm e^\pm \mu^\mp \tau^\mp \,,
\end{eqnarray}
which is more challenging, as a result of the small branching fraction of $H^{\pm\pm} \to \mu^\pm \tau^\pm$ and the low $\tau$ efficiency~\cite{CMS:2017pet}.

\subsection{Prospects at lepton colliders}

\begin{figure}[!t]
  \includegraphics[width=0.28\textwidth]{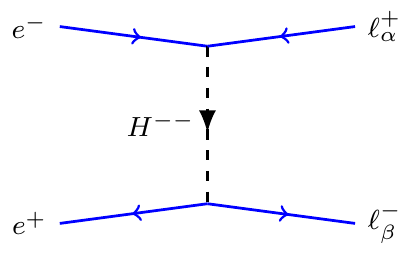}
  \caption{LFV signal from an off-shell doubly-charged scalar at lepton colliders.}
  \label{fig:diagram}
\end{figure}

\begin{figure}[!t]
  \centering
  \includegraphics[width=0.48\textwidth]{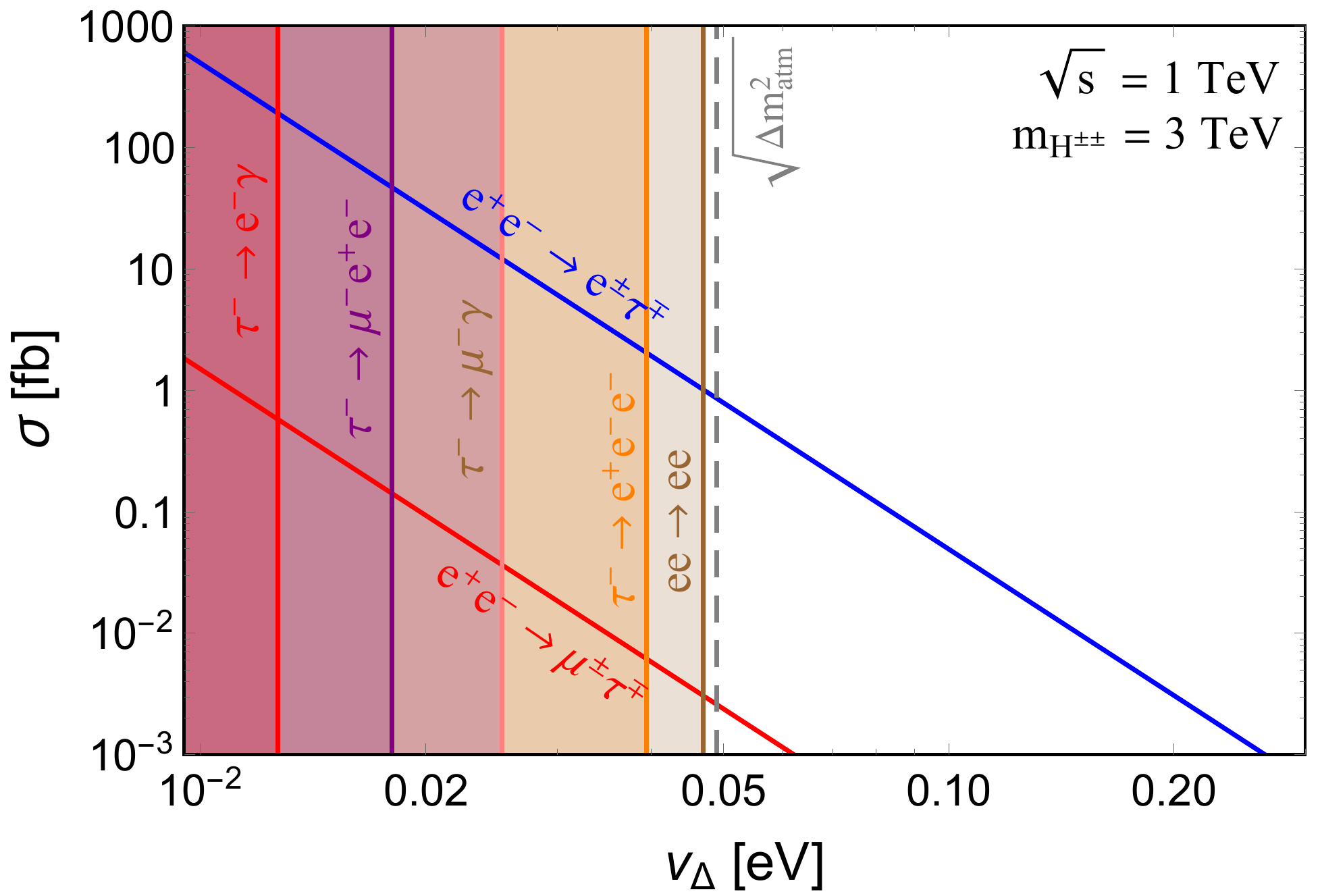}
  \caption{Cross section of the LFV process $e^+ e^- \to e^\pm \tau^\mp$, $\mu^\pm \tau^\mp$ at ILC at $\sqrt{s} = 1$ TeV with polarized beam of $P_{e-} = -0.8$ and $P_{e^+} = +0.3$, as function of the VEV $v_\Delta$ with a massless lightest neutrino ($m_3 = 0$) in the case of inverted hierarchy. Also shown are the constraints from the rare $\tau$ decays and $ee \to ee$ data at LEP, as indicated. The vertical dashed line corresponds to a mass $\sqrt{\Delta m^2_{\rm atm}} \simeq$ 0.05 eV.}
  \label{fig:offshell}
\end{figure}

Even though a 3 TeV doubly-charged scalar can not be directly (pair) produced at future lepton colliders like ILC at $\sqrt{s} = 1$ TeV, it could induce off-shell LFV violating signals via the diagram in Fig.~\ref{fig:diagram}~\cite{Dev:2017ftk}. The couplings of doubly-charged scalar to the charged leptons are totally determined by the neutrino mass matrix~(\ref{eq:neutrino}), rescaled by the VEV $v_\Delta$. To be concrete, we assume the active neutrino are of inverted hierarchy, as implied by the DAMPE data, and $m_3 = 0$. With $m_{H^{\pm\pm}} = 3$ TeV, one is ready to obtain the LFV cross section $\sigma (e^+ e^- \to \ell_i^\pm \ell_j^\mp)$ as functions of the VEV $v_\Delta$. The expected cross sections at ILC are presented in Fig.~\ref{fig:offshell},\footnote{It is also feasible to search for the doubly-charged scalar induced LFV signals at CEPC~\cite{CEPC-SPPCStudyGroup:2015csa} and FCC-ee~\cite{Gomez-Ceballos:2013zzn}, which is however less promising, as a result of the lower colliding energy.} for the two cases of $e\tau$ and $\mu\tau$ (the $e\mu$ channel is severely constrained by the rare $\mu$ decay data), with $\sqrt{s} = 1$ TeV and polarized beams of $P_{e^-} = -0.8$ and $P_{e^-} = +0.3$, which enhances the cross sections by a factor of $(1- P_{e^-}) (1+ P_{e^+})  = 2.34$. We have applied the nominal cuts of $p_T > 10$ GeV on the charged leptons and take an efficiency factor of 60\% for the tau lepton~\cite{Baer:2013cma}. Note that the cross sections have a fourth power dependence on the VEV, i.e. $\sigma \propto v_{\Delta}^{-4}$, while the amplitude in Fig.~\ref{fig:diagram} have a quadratic dependence on the lepton flavor conserving/violating couplings $(Y_\Delta)_{ij}$.


Note that the LFV couplings in Eq.~(\ref{lag}) induce also rare flavor violating decays and anomalous magnetic moments which are highly suppressed in the SM.  The contribution of doubly-charged scalar loops to the electron $g-2$ is~\cite{Leveille:1977rc, Moore:1984eg, Gunion:1989in}
\begin{eqnarray}
&&\Delta a_e = - \sum_\ell
\frac{m_e^2 (Y_\Delta)_{e\ell}^2}{12\pi^2 m_{H^{\pm\pm}}^2} \,.
\end{eqnarray}
To set limits on the VEV $v_\Delta \propto (Y_\Delta)^{-1}$, we set the lightest neutrino mass $m_3 = 0$ in the inverted hierarchy and summed up the loops involving all the three flavors $\ell = e,\, \mu,\, \tau$. The current $2\sigma$ experimental uncertainty $\Delta a_e = 5.2 \times 10^{-13}$~\cite{PDG} imposes a lower bond on the VEV $v_\Delta > 0.0011$ eV. In an analogous way, one can calculate the contribution to the muon $g-2$. As the contributions from the doubly-charged loops are always negative, the controversial theoretical and experimental discrepancy $\Delta a_\mu = (2.87 \pm 0.80) \times 10^{-9}$ can not be explained. We use instead the $5\sigma$ uncertainty to constrain $v_\Delta$, which is stronger than the electron $g-2$ and requires that $v_\Delta > 0.002$ eV.

The partial width for the rare LFV decay $\mu \to eee$ is~\cite{Akeroyd:2009nu, Dinh:2012bp}
\begin{eqnarray}
\Gamma (\mu \to eee) \simeq \frac{|(Y_\Delta)_{ee}|^2 |(Y_\Delta)_{e\mu}|^2}{16 G_F^2 m_{H^{\pm\pm}}^4} {\rm BR} (\mu \to e\nu\bar\nu) \,,
\end{eqnarray}
with $G_F$ the Fermi constant. Given $m_3 = 0$, the current experimental data BR($\mu \to eee$)$< 10^{-12}$~\cite{PDG} exclude $v_\Delta$ up to 0.78 eV, and there is no hope to see any $e^+ e^- \to e^\pm \mu^\mp$ events at ILC down to the cross section of $10^{-3}$ fb (cf. the $e\tau$ line in Fig.~\ref{fig:offshell}). Similarly, the data ${\rm BR} (\tau^- \to e^+ e^- e^-) < 2.7 \times 10^{-8}$ and ${\rm BR} (\tau^- \to \mu^- e^+ e^-) < 1.8 \times 10^{-8}$~\cite{PDG} can be used to constrain $|(Y_\Delta)_{ee}^\ast (Y_\Delta)_{e\tau}|$ and $|(Y_\Delta)_{e\mu}^\ast (Y_\Delta)_{e\tau}|$, or effectively on $v_\Delta$ as shown in Fig.~\ref{fig:offshell}, which are respectively relevant  to the production of $e\tau$ and $\mu\tau$.

At 1-loop level we have the two-body LFV decays~\cite{Mohapatra:1992uu,Akeroyd:2009nu, Dinh:2012bp}
\begin{eqnarray}
{\rm BR} (\ell_i \to \ell_j \gamma) & \ \simeq \ &
\frac{\alpha |\sum_k (Y_\Delta)_{ik}^\dagger (Y_\Delta)_{jk}|^2}{12\pi G_F^2 M_{\pm\pm}^4} \nonumber \\ &&
 \times {\rm BR} (\ell_i \to e\nu\bar\nu) \,,
\end{eqnarray}
where $\alpha$ is the fine structure constant, and we have summed up all the diagrams involving a $\ell_k$ lepton running in the loop. In the type-II seesaw, it is equivalent to do the summation $\sum_k (m_\nu)_{ik}^{\sf T} (m_\nu)_{jk}$. The experimental data of ${\rm BR} (\mu \to e \gamma) < 4.2 \times 10^{-13}$~\cite{PDG} implies that the VEV $v_\Delta > 0.32$ eV, and preclude the signal of $ee \to e\mu$ at ILC. The current limits ${\rm BR} (\tau \to e\gamma) < 3.3 \times 10^{-8}$ and ${\rm BR} (\tau \to \mu \gamma) < 4.4 \times 10^{-8}$ could be used to set limits on the couplings $|\sum_k (Y_\Delta)_{\tau k}^\dagger (Y_\Delta)_{e k}|$ and $|\sum_k (Y_\Delta)_{\tau k}^\dagger (Y_\Delta)_{\mu k}|$, and the constraints on the VEV $v_\Delta$ for a 3 TeV doubly-charged scalar are comparable to those from the three-body LFV decays $\tau \to \ell_i \ell_j \ell_k$ above, as shown in Fig.~\ref{fig:offshell}.

The doubly-charged scalar could also mediate the scattering $ee \to ee,\, \mu\mu,\, \tau\tau$ at LEP (cf. Fig.~\ref{fig:diagram}), which would interfere with the SM background. Both the total cross section and differential distributions would be modified by the presence of beyond SM couplings $(Y_\Delta)_{ij}$. Benefiting from the larger coupling to the electron flavor for the inverted hierarchy, the most stringent limit is from $ee \to ee$, which excludes an effective cutoff scale of $\Lambda \simeq m_{H^{\pm\pm}}/|(Y_\Delta)_{ee}| < 5.3$ TeV~\cite{Abdallah:2005ph}. When applied to the type-II seesaw, it is required that  $v_\Delta > 0.047$ eV in the limit of $m_3 = 0$, as shown in Fig.~\ref{fig:offshell}.

It is worth mentioning that future higher energy lepton colliders like ILC could improve largely the LEP limits above such as in the process $e^+ e^- \to e^+ e^-$. At $\sqrt{s} = 500$ GeV and with an integrated luminosity of 500 fb$^{-1}$, the effective cutoff scale $\Lambda$ could pushed up to 82 TeV~\cite{Pankov:2005kd, Pankov:2007ey} (with a higher $\sqrt{s}$ and larger luminosity the reaches could be higher) and exclude the coupling $(Y_\Delta)_{ee}$ larger than 0.036 for a 3 TeV doubly-charged scalar, which corresponds to the value of $v_\Delta = 1.3$ eV for the IH case. It is clear that the all the parameter space region for the DAMPE anomaly and the LFV signals $ee \to \ell_i \ell_j$ in Fig.~\ref{fig:offshell} can be directly tested at future lepton colliders.

The singly-charged scalar would be single produced at lepton colliders through $e^\pm \gamma \to \nu_i H^{\pm \, (\ast)}$, besides the Drell-Yan process, which is however much less promising than the off-shell doubly-charged scalar if $H^\pm$ is too heavy, e.g. 3 TeV, to be produced on-shell. In addition, a singly-charged scalar induces the scattering of astrophysical and atmospheric neutrinos off electrons in the IceCube detector, and produces charged leptons $\ell_i$ of all the three flavors, i.e. $\nu_i e \to \ell_j \nu_k$. This interferes with the SM processes mediated by the $W$ boson, and contributes to the effective area at IceCube for both the neutrinos and antineutrinos of all the flavors~\cite{Aartsen:2013jdh, Aartsen:2014muf, Aartsen:2017mau}, with roughly an enhancement factor of
\begin{eqnarray}
\sim \left[ 1 + \sum_{ijk} \left( \frac{|(Y_\Delta)_{ij}^\ast (Y_\Delta)_{ek}|}{2g^2} \right)
\left( \frac{m_W^2}{m_{H^\pm}^2} \right) \right] \,,
\end{eqnarray}
where we have summed up all the flavors for the incoming neutrinos and outgoing neutrinos and charged leptons. Then the IceCube observed data and its $1\sigma$ errors can be used to set limits on the couplings $Y_\Delta$ and the VEV $v_\Delta$~\cite{Aartsen:2016xlq}. It turns out that the constraint is rather loose, $v_\Delta \gtrsim 0.006$ eV, and is not shown in Fig.~\ref{fig:offshell}.

Given all the limits above, there is still large parameters space in Fig.~\ref{fig:offshell} unconstrained, and the cross section $ee \to e\tau$ and $ee \to \mu\tau$ could reach up to 1.0 fb and 0.003 fb, respectively. It is very promising that the minimal type-II seesaw could be directly tested at ILC, in particular in the $e\tau$ channel, although the 3 TeV scalars can not be directly (pair) produced on-shell. At CLIC with a higher center-of-mass energy at 3 TeV, the cross sections are expected to be much larger.

\section{Discussions and conclusion}
\label{sec:conclusion}

In this paper we have pointed out that the tentative DAMPE $e^+ e^-$ peak at 1.4 TeV over the power law background can be understood in terms of the type-II seesaw model with a scalar DM. if both the DM $D$ and triplet scalars have a mass of $\sim3$ TeV. With the mass ordering $m_{H^{\pm\pm}} < m_D < m_{H^\pm} < m_{H,A}$, the $e^+ e^-$ excess can be obtained via the DM annihilation $DD \to H^{++} H^{--}$ and the subsequent decay $H^\pm \to e^\pm e^\pm$. This is the simplest explanation in the framework of type-II seesaw model. 
The secondary neutrinos and photons from $H^{\pm\pm} \to \ell_i^\pm \ell_j^\pm$ ($i,\,j = \mu,\,\tau$) are much softer and orders of magnitude below the current observations.

If the triplet scalars are around 3 TeV and leptophilic, then all the neutral, singly and doubly-charged components can only be produced on-shell at future 100 TeV colliders, e.g. by searching for the lepton number and flavor violating signals like $(e^+ e^+) (e^- e^-)$ and $(ee)(\mu\tau)$. An alternative way is to produce the doubly-charged scalars off-shell at future lepton colliders, which mediates LFV signals like $e^+ e^- \to e^\pm \tau^\mp$. A broad parameter region of type-II seesaw can be probed, with a cross section up to 1 fb at ILC, which is still allowed by the current stringent low energy lepton flavor constraints (see Fig.~\ref{fig:offshell}).

If the triplet scalars are significantly lower than 3 TeV, say 1 TeV or even lower, all the analysis here holds true and the future hadron and lepton collider searches are expected to be much more promising, and largely complementary to the high intensity frontier experiments and the neutrino experiments. This will be pursued in a more generic sense in a upcoming publication~\cite{nextpaper}.

\section*{Acknowledgements}

The authors would like to thank Rabindra N. Mohapatra and P. S. Bhupal Dev for the enlightening discussions throughout the whole process of this paper. We are also grateful to Marco Drewes for the valuable comments. Y.Z. would like to thank the Center for High Energy Physics, Peking University for the hospitality and local support on the visit.

\end{document}